\documentclass[draft]{agujournal2019}
\usepackage{url} 
\usepackage[inline]{trackchanges} 
\usepackage{soul}
\usepackage{natbib}
\usepackage{amssymb}

\draftfalse

\journalname{Enter journal name here}

\begin{document}

%
%


\title{Recovery of Directional Wave Spectrum from Sparse Data with Compressed Sensing}

%
%





\authors{
Qingyu Jiang\affil{1}, Henrik Kalisch\affil{1}, Michel Benoit\affil{2,3}, Karoline Holand\affil{1}, Patrick Sprenger\affil{4}
}


\affiliation{1}{Department of Mathematics, University of Bergen, Bergen, Norway}
\affiliation{2}{EDF R\&D, Laboratoire National d'Hydraulique et Environnement (LNHE), Chatou, France}
\affiliation{3}{LHSV, Saint-Venant Hydraulics Laboratory, ENPC, Institut Polytechnique de Paris, Chatou, France}
\affiliation{4}{Department of Applied Mathematics, University of California, Merced, USA}

\correspondingauthor{Henrik Kalisch}{henrik.kalisch@uib.no}

\begin{abstract}
[ 
Compressed sensing provides an efficient framework for reconstructing wave signals from reduced measurements. 
For multi-channel buoy data, the three displacement components exhibit intrinsic correlations, 
as wave motion contributes simultaneously to all directions according to linear wave theory. 
Meanwhile, conventional compressed sensing methods based on $\ell_1$-shrinkage tend to underestimate 
signal energy when sparsity is not strictly satisfied, leading to biased spectral estimation. 
This paper introduces a group sparsity constraint to promote physically consistent sparse 
representations across channels. An energy constraint is proposed in the form of a soft lower bound, 
enabling an isotropic rescaling of the recovered spectrum while preserving its sparse structure. 
Considering a large volume of buoy data, we demonstrate through a series of experiments 
that the proposed approach enables compression by retaining a subset of original measurements.
]
\end{abstract}

%
%

\section{Introduction}
Directional wave spectra provide detailed information on the distribution of wave energy across frequencies and directions, 
and are widely used for ocean analysis and engineering applications \citep{holthuijsen2007waves}. 
In practice, reliable spectral estimation typically requires sufficient data to enable averaging 
and achieve statistical stability \citep{welch1967use, thomson1982spectrum}, 
and buoy measurements are commonly processed over extended time windows \citep{tucker2001waves}. 
As a result, high-resolution wave spectral analysis is often associated with large volumes of data. 
However, ocean wave signals exhibit structured and correlated behavior, 
and the spectral representation itself is obtained through statistical 
averaging rather than instantaneous measurements \citep{longuet1963random}. 
This suggests that full-resolution sampling of the original time series may not be strictly necessary 
to capture the essential spectral characteristics. It is therefore of interest to investigate whether 
wave spectra can be accurately estimated from a reduced set of measurements without requiring complete data acquisition.\\

Compressed sensing provides a natural framework for recovering signals from reduced measurements 
as noted in \citep{donoho2006compressed}, 
and has been widely studied for signal reconstruction \citep{duarte2011structured}. 
The standard Compressed sensing framework relies on the assumption that the signal admits a sparse representation 
when represented in appropriate variable, which is often satisfied for ocean wave data, 
but not always in the strict sense required by the standard compressed sensing approach \citep{candes2008introduction}. 
While wave spectra often exhibit energy concentration around dominant components \citep{holthuijsen2007waves}, 
spectral spreading may exist due to the coexistence of swell, wind-sea and nonlinear interactions \citep{hasselmann1962non}. 
Natural signals are usually not strictly sparse but often compressible in the transform domain.
In compressed sensing terminology such as used in \citep{donoho2006compressed} and \citep{candes2008introduction}, 
wave spectra could be characterized as data-compressible rather than strictly sparse 
while still retaining sufficient structure for reconstruction.\\

Another benefit of compressed sensing is the possibility to reduce the data storage requirements for onboard recording systems in wave buoys. 
Modern wave buoys typically employ on‑board data logging to complement or back up real-time telemetry, 
and to preserve high-resolution measurements that cannot be transmitted continuously due to bandwidth and power limits. 
In most designs, sensor data such as raw acceleration, heave–pitch–roll motions, and diagnostic parameters are written 
to solid-state memory. This local logging allows the buoy to retain complete time series during communications outages 
and enables detailed post-recovery analysis that goes beyond transmitted summary statistics or spectra. 
For example, the Datawell Directional Waverider stores raw motion data internally while transmitting processed 
wave parameters via radio or satellite. In continuous operation, wave records must be kept over periods 
of several months to years. If the recorded data can be compressed before storage, 
then the deplpoyment period could be extended accordingly. \\ 

This paper develops a theoretical compressed sensing framework for multi-channel wave data. 
The formulation incorporates a structured sparsity model to capture the joint behavior of displacement components, 
enabling a more consistent representation across channels. We also use energy-related regularization to improve 
the stability of the reconstruction and enhance the fidelity of the estimated spectrum. 
Then we conduct experiments through Bideford Bay buoy data \citep{dataset_NNRCMP}, 
and the performance shows our approach allows wave spectrum reconstruction from reduced measurements 
while maintaining essential spectral features and can be extended to cases with incomplete observations.\\

The contributions of the paper are the following:
\begin{enumerate}
    \item  We develop a unified physics-informed compressed sensing framework for multi-channel wave reconstruction that simultaneously addresses cross-channel structural inconsistency and systematic energy bias of conventional $\ell_1$-based recovery.
    \item  We derive an  Alternating Direction Method of Multipliers (ADMM)-based optimization strategy that separates sparse structure identification from energy rescaling, making the reconstruction more stable and interpretable.
    \item We provide an experimental analysis of recoverability under different spectral conditions, showing that reconstruction quality is controlled by the interplay between sparsity and sampling ratio.
    \item We further demonstrate that directional spectrum estimation remains usable after compressed sensing reconstruction, while also revealing its structural limitations.
\end{enumerate}

The remainder of this article is organized as follows: in Section \ref{sec:method}, we review and discuss the principles of the compressed sensing theory and its practical implementation. Experiments and application of the method with real wave buoy data, are presented in Section \ref{sec:expes}. The capabilities and limitations of the compressed sensing are assessed and discussed in Section \ref{sec:discus}, while Section \ref{sec:conclusion} summarizes the main findings of this study.

\section{Method} \label{sec:method}

\noindent
\subsection{Compressed Sensing}
Compressed sensing recovers signals from reduced measurements by exploiting sparse representations 
in an appropriate basis. Let $\mathbf{x}\in\mathbb{R}^{N\times M}$ denote the multi-channel wave signal, 
where $N$ is the number of temporal samples and $M$ is the number of channels. 
The observed data $\mathbf{y}\in\mathbb{R}^{N\times M}$ are assumed to be partially sampled and can be written as 

\begin{equation}
    \mathbf{y} = \mathbf{C}\mathbf{x}
\end{equation}
where $C\in\mathbb{R}^{N\times M}$ is a sampling matrix, typically diagonal with binary entries indicating observed locations.

To exploit the spectral structure of wave signals, the signal is represented in a Fourier basis
\begin{equation}
    \mathbf{x} = \Psi\mathbf{s}
\end{equation}
where $\Psi\in\mathbb{C}^{N\times N}$ is the Fourier basis matrix and $\mathbf{s}\in\mathbb{C}^{N\times M}$ 
denotes the corresponding spectral coefficients. The Fourier basis provides a direct representation of spectral content, 
since the energy distribution of a signal across frequencies is defined via its Fourier transform \citep{welch1967use}. 
Compared with overcomplete dictionaries, orthonormal bases such as the Fourier basis provide simpler structure 
and are often associated with more favorable incoherence properties in compressed sensing frameworks \citep{elad2010sparse}.

\subsubsection{Restricted Isometry Properties (RIP)}
In compressed sensing, stable recovery is associated with low coherence \citep{candes2005decoding}, which requires
the estimate
\begin{equation}
    (1 - \delta_k)\,\|\mathbf{s}\|_2^2 \le \|\mathbf{C}\Psi \mathbf{s}\|_2^2 \le (1 + \delta_k)\,\|\mathbf{s}\|_2^2
\end{equation}
for small $\delta_k > 0$.
A signal is said to be k-sparse if it is supported on at most k coefficients in a given basis, 
following the standard compressed sensing formulation \citep{candes2008introduction}. 
To reduce coherence, $\mathbf{C}$ is constructed via random sampling.

For wave spectra in our model, the coefficient vector $\mathbf{s}$ is not strictly sparse but compressible. 
As a result, the above RIP condition is only approximately satisfied, 
and the energy preservation implied by compressed sensing is weakened.

\subsection{Model formulation and ADMM optimization}
We propose an optimization model to recover a physically consistent spectral representation from incomplete measurements 
while compensating for the energy loss introduced by sparse regularization. 
The problem is solved using the Alternating Direction Method of Multipliers (ADMM) framework \citep{boyd2011distributed}. 
The elementwise and group sparsity terms are used to identify the dominant spectral structure, 
while the lower-bound energy term corrects the systematic underestimation of signal energy. 
The original sparse reconstruction problem is convex, but the introduction of an energy constraint 
makes the overall problem no longer fully convex. Thus, we adopt the ADMM framework to split the optimization 
into two coupled subproblems to preserve the convex structure as much as possible, 
while the energy correction is handled separately. ADMM requires each subproblem to be solved to optimality. 
In our implementation, the ${\mathbf{s}}$-step is optimized via a proximal gradient method until convergence, 
and then alternated with the ${\mathbf{z}}$-step to obtain a stable solution. \\

We consider the spectral coefficient matrix $\mathbf{S}\in\mathbb{C}^{N\times M}$, 
where $N$ denotes the number of frequency indices and $M$ denotes the number of channels. 
Each entry $\mathbf{s}_{i,j}$ represents the coefficient at frequency index $i$ and channel $j$.
Introducing an auxiliary variable $\mathbf{z}\in\mathbb{C}^{N\times M}$, 
the problem can be rewritten as
\begin{equation}
\min_{\mathbf{s},\, \mathbf{z}} \;
\frac{1}{2}\,\|\mathbf{C}\Psi \mathbf{s} - \mathbf{y}\|_F^2
+ \lambda_e \|\mathbf{s}\|_1
+ \lambda_g \sum_{i=1}^{N} \|\mathbf{s}_{i,:}\|_2
+ \mu \sum_{j=1}^{M} \left( \max\{0,\, E_j - \|\mathbf{s}\|_2^2 \} \right)^2,
\end{equation}
where $\|\cdot\|_F$ indicates the Frobenius norm.
The corresponding augmented Lagrangian is
\begin{equation}
    \mathcal{L}(\mathbf{s}, \mathbf{z}, \mathbf{u}) =
\frac{1}{2} \| \mathbf{C}\Psi \mathbf{s} - \mathbf{y} \|_F^2
+ \lambda_e \| \mathbf{s} \|_1
+ \lambda_g \sum_{i=1}^{N} \| \mathbf{s}_{i,:} \|_2
+ \mu \sum_{j=1}^{M} \left( \max \{ 0,\, E_j - \| \mathbf{z}_{:,j} \|_2^2 \} \right)^2
+ \frac{\rho}{2} \| \mathbf{s} - \mathbf{z} + \mathbf{u} \|_F^2 
\end{equation}
so that $\mathbf{s}  = \mathbf{z}$, 
where $\mathbf{u}\in\mathbb{C}^{N\times M}$ is the scaled dual variable and $\rho >0$ is the penalty parameter.\\

Some rigorous methods exist for bias-free spectral estimation from data with missing samples \citep{damaschke2024bias}. 
Here, we adopt a simpler approximation based on rescaling the observed energy. 
The quantity $E_j$ represents the estimated total energy of the $j$-th channel 
computed from the observed samples and used as a target level in the energy regularization term. 
The number of observed samples is given by $M = \sum_{i,j} C_{i,j}$, 
and the observed energy is computed as $E_{\mbox{obs},j} = \sum_{i=1}^{N} \left| \mathbf{y}_{i,j} \right|^2$.\\

Assuming that the observed samples are representative of the full signal, 
the total energy in each channel is estimated by rescaling the average observed squared magnitude 
to the full signal length: $E_j = e_r \frac{N}{M} E_{\mbox{obs},j}$, 
where $e_r$ is a scaling parameter controlling the conservativeness of the energy estimate. 
It is independent of the parameter $\mu$ in the optimization model 
which appears in the energy regularization term and determines the strength of the energy constraint during reconstruction.

\subsubsection{Sparse reconstruction subproblem} 
With $\mathbf{z}$ and $\mathbf{u}$ fixed, the update of $\mathbf{s}$ is obtained from:
\begin{equation}
\mathbf{s}^{k+1} = \arg\min_{\mathbf{s}} \;
\frac{1}{2} \| \mathbf{C}\Psi \mathbf{s} - \mathbf{y} \|_F^2
+ \lambda_e \| \mathbf{s} \|_1
+ \lambda_g \sum_{i=1}^{N} \| \mathbf{s}_{i,:} \|_2
+ \frac{\rho}{2} \| \mathbf{s} - \mathbf{z}^k + \mathbf{u}^k \|_F^2 .
\end{equation}

This subproblem retains the structure of the original convex sparse reconstruction
as the nonconvex energy term is removed at this stage. The first term enforces data consistency, 
the second term promotes element-wise sparsity controlled by $\lambda_e$, 
and the third term corresponds to the mixed $\ell_{2,1}$-norm 
which enforces group sparsity across frequency indices, with its strength governed by $\lambda_g$
which is consistent with the shared sparsity profile assumption 
in multi-measurement vector problems \citep{cotter2005sparse}. 
Specifically, it encourages the coefficients $\mathbf{s}_i$ at each frequency to be jointly selected across channels. \\

Both parameters $\lambda_e$ and $\lambda_g$ act on the shrinkage intensity 
rather than the update step size, and after normalization, 
they can be treated as dimensionless coefficients selected empirically to balance sparsity, 
noise suppression, and reconstruction accuracy.\\

The last term corresponds to the quadratic penalty in the augmented objective
which enforces consistency between the primary variable ${\mathbf{s}}$ 
and the auxiliary variable ${\mathbf{z}}$. The parameter $\rho$ controls the strength of this coupling. 
Under normalization, $\rho$ operates on the same scale as the data fidelity 
term and is therefore comparable to $\lambda_e$ and $\lambda_g$ in magnitude.\\

To solve this subproblem, a proximal gradient scheme is employed \citep{beck2009fast} 
which is suitable for composite optimization problems consisting of a smooth term and a non-smooth regularization. 
Define the smooth part
\begin{equation}
f(\mathbf{s}) = \frac{1}{2} \| \mathbf{C}\Psi \mathbf{s} - \mathbf{y} \|_F^2
+ \frac{\rho}{2} \| \mathbf{s} - \mathbf{z}^k + \mathbf{u}^k \|_F^2 .
\end{equation}
whose gradient is

\begin{equation}
\nabla f(\mathbf{s}) = \Psi^H \mathbf{C}^T \bigl( \mathbf{C}\Phi \mathbf{s} - \mathbf{y} \bigr)
+ \rho \bigl( \mathbf{s} - \mathbf{z}^k + \mathbf{u}^k \bigr).
\end{equation}
Starting from $s^k$, then apply a gradient descent step:
\begin{equation}
    \tilde{s}_{i,j}\leftarrow s^k_{i,j}-\tau\nabla f({\mathbf{s}}^k_{i,j}),
\end{equation}
where $\tau >0$ is the step size, chosen as $\tau = \frac{1}{1+\rho}$
since the augmented Lagrangian term increases the curvature, 
resulting in an effective Lipschitz constant of  $1+\rho$ under an orthonormal Fourier basis.\\

The updated variable $\mathbf{s}$ is then processed by two successive shrinkage operations.\\
For the element-wise sparsity term, each coefficient is updated as
\begin{equation}
\overline{{\mathbf{s}}}_{i,j} 
= \frac{\tilde{\mathbf{s}}_{i,j}}{|\tilde{s}_{i,j}|}\cdot \max \bigl( |\tilde{\mathbf{s}}_{i,j}|-\tau\lambda_e, 0\bigr)
\end{equation}
with the convention that the result is zero when $\mathbf{s}_{i,j}=0$.\\
For the group sparsity term, the update is applied row-wise. Let
\begin{equation}
    \|\mathbf{s}_{i,:}\|_2 = \sqrt{\sum^M_{j=1} \mathbf{s}_{i,j}^2}
\end{equation}
then the entire row is updated as

\begin{equation}
    \mathbf{s}_{i,j} = \max \bigl( 1- \frac{\tau\lambda_g}{\|\overline{s}_{i,:}\|_2}, 0\bigr)\cdot \mathbf{s}_{i,j}.
\end{equation}
These two shrinkage operations are applied sequentially. 
The element-wise shrinkage identifies a sparse set of active coefficients
while the group-wise shrinkage enforces a shared frequency support across channels. 
In this way, the main spectral structure is determined in this stage before the energy correction is applied.
\subsubsection{Energy correction subproblem}
With $\mathbf{s}$ and $\mathbf{u}$ fixed, the update of $z$ is obtained from

\begin{equation}
\mathbf{z}^{k+1} = \arg\min_{\mathbf{z}} \;
\mu \sum_{j=1}^{M} \left( \max \{ 0,\, E_j - \| \mathbf{z}_{:,j} \|_2^2 \} \right)^2
+ \frac{\rho}{2} \| \mathbf{z} - (\mathbf{s}^{k+1} + \mathbf{u}^k) \|_F^2 
\end{equation}
This problem is separable across channels. 
For each channel $j$, define $\mathbf{q}_{i,j} = \mathbf{s}_{i,j}^{k+1} + \mathbf{u}_{i,j}^k$. 
Then the subproblem becomes
\begin{equation}
\min_{\mathbf{z}} \;
\mu \left( \max \{ 0,\, E_j - \| \mathbf{z}_{:,j} \|_2^2 \} \right)^2
+ \frac{\rho}{2} \| \mathbf{z}_{:,j} - \mathbf{q}_{:,j} \|_2^2 
\end{equation}
This objective depends only on the norm of $\mathbf{z}_{:,j}$, and is therefore isotropic. 
As a result, the solution must lie along the direction of $\mathbf{q}_{:,j}$, and can be written as
\begin{equation}
    \mathbf{z}_{i,j} = \frac{r}{\| \mathbf{q}_{:,j} \|_2} \, \mathbf{q}_{i,j}, \quad r \ge 0
\end{equation}

Let $a = \|\mathbf{q}_{:,j}\|_2$. Substituting this form reduces the problem to scalar optimization:

\begin{equation}
\min_{r \ge 0} \;
\mu \left( \max \{ 0,\, E_j - r^2 \} \right)^2 + \frac{\rho}{2} (r - a)^2 
\end{equation}
If $a^2\ge E_j$, the energy constraint is inactive and the solution is 
$$
r = a, \quad \mathbf{z}_{i,j} = \mathbf{q}_{i,j}.
$$
If $a^2< E_j$, the objective becomes $$\mu (E_j - r^2)^2 + \frac{\rho}{2} (r - a)^2.$$
Differentiation with respect to $r$ leads to the cubic equation $$4\mu r^3 + (\rho - 4\mu E_j)\, r - \rho a = 0.$$
The optimal radius $r$ is obtained from this equation, and the update is the same as before
\begin{equation}
    \mathbf{z}_{i,j} = \frac{r}{\| \mathbf{q}_{:,j} \|_2} \, \mathbf{q}_{i,j}
\end{equation}

This form shows that the energy correction does not change the direction of the coefficient vector, and only rescales its magnitude. Therefore, the spectral structure obtained from the sparse reconstruction step is preserved, and the energy is adjusted without destroying the sparsity pattern.

\subsubsection{Dual update}
Finally, the dual variable is updated as
\begin{equation}
    \mathbf{u}^{k+1} = \mathbf{u}^k + \mathbf{s}^{k+1} - \mathbf{z}^{k+1}.
\end{equation}
The iterations are repeated until the error drops below a specified numerical tolerance.

\subsection{Spectral Estimation}
To evaluate the reconstructed signals in terms of directional wave properties, 
the directional spectrum is estimated from the three displacement components. 
Following standard theory \citep{holthuijsen2007waves}, the directional spectrum is expressed as
\begin{equation}
    S(f, \theta) = E(f)\, D(f, \theta),
\end{equation}
where $E(f)$ is the one-dimensional frequency spectrum 
and $D(f, \theta)$ is the directional spreading function, satisfying
\begin{equation}
    \int_{0}^{2\pi} D(f,\theta)\, d\theta = 1.
\end{equation}
For single-point buoy measurements, the three displacement components are used 
to construct the cross-spectral matrix $\mathbf{G}(f)$, whose elements are defined as
\begin{equation}
   G_{mn}(f) = \left\langle X_m(f)\, X_n^*(f) \right\rangle,
\end{equation}
where $X_m(f)$ denotes the Fourier transform of the $m$-th signal component 
and $*$ denotes the complex conjugation.\\
The cross-spectral matrix is related to the directional spectrum through
\begin{equation}
    G_{mn}(f) = \int_{0}^{2\pi} H_m(f,\theta)\, H_n^*(f,\theta)\, S(f,\theta)\, d\theta,
\end{equation}
where $H_m(f, \theta)$ are the transfer functions derived from linear wave theory, 
relating the wave elevation to the measured components. Since only a finite number 
of cross-spectral quantities are available, the estimation of $D(f, \theta)$ from $\mathbf{G}(f)$ 
constitutes an underdetermined inverse problem and therefore requires additional assumptions or regularization.

\subsubsection{Iterative refinements of MLM - version 2 (IMLM2)}
The Maximum Likelihood Method (MLM) models the directional distribution as a linear combination of cross-spectra, 
leading to an estimate that can be interpreted as a convolution with a window function approaching 
a Dirac delta under ideal conditions, while the IMLM2 method iteratively refines this estimate to enforce 
the consistency with the measured cross-spectra \citep{krogstad1988mlm, benoit1997comparative}.\\

The directional spreading function can be initially estimated using the Maximum Likelihood Method (MLM), defined as
\begin{equation}
    \hat{D}_{\mathrm{MLM}}(f,\theta)
= \kappa \left( \mathbf{H}^H(f,\theta)\, \mathbf{G}^{-1}(f)\, \mathbf{H}(f,\theta) \right)^{-1},
\end{equation}
where $\mathbf{G}(f)$ is the cross-spectral matrix, $\mathbf{G}^{-1}(f)$ its inverse.
In this formula,
$\mathbf{H}(f,\theta) = \left[ H_1(f,\theta),\, H_2(f,\theta),\, H_3(f,\theta) \right]^T$ 
is the transfer function vector, $(\cdot)^H$ denotes the Hermitian transpose, and $\kappa$ is a normalization constant ensuring 
\begin{equation}
\int_{0}^{2\pi} \hat{D}_{\mathrm{MLM}}(f,\theta)\, d\theta = 1.
\end{equation}
The IMLM method refines this estimate iteratively \citep{benoit1997comparative}
using the relation
\begin{equation}
\hat{D}^{(i)}_{\mathrm{IMLM}}(f,\theta)
= \hat{D}^{(i-1)}_{\mathrm{IMLM}}(f,\theta)
+ \varepsilon^{(i)}(f,\theta),
\quad
\hat{D}^{(0)}_{\mathrm{IMLM}}(f,\theta)
= \hat{D}_{\mathrm{MLM}}(f,\theta).
\end{equation}
For IMLM2, the correction term is defined as
\begin{equation}
\varepsilon^{(i)}(f,\theta)
= \frac{\left| \lambda^{(i)}(f,\theta) \right|^{\beta+1}}{\gamma},
\quad
\lambda^{(i)}(f,\theta)
= \hat{D}_{\mathrm{MLM}}(f,\theta)
- \hat{D}^{(i-1)}_{\mathrm{MLM}}(f,\theta),
\end{equation}
where $\hat{D}^{(i-1)}_{\mathrm{MLM}}(f,\theta)$ denotes the MLM estimate recomputed 
from the cross-spectral matrix associated with the previous iteration.

\subsubsection{Maximum Entropy Method - version 2 (MEM2)}
The MEM2 method determines the directional distribution by maximizing entropy under moment constraints, 
resulting in an exponential-form solution that enforces non-negativity, preserves only the information 
contained in the measured moments \citep{ benoit1999comparative}.
It estimates the directional spreading function by maximizing the Shannon entropy
\begin{equation}
    H = - \int_{0}^{2\pi} D(f,\theta)\, \ln D(f,\theta)\, d\theta
\end{equation}
subject to constraints imposed by the cross-spectral coefficients. The resulting solution has the form
\begin{equation}
    D(f,\theta) = \exp\bigl(
- \lambda_0(f)
- \lambda_1(f)\cos\theta
- \lambda_2(f)\sin\theta
- \lambda_3(f)\cos 2\theta
- \lambda_4(f)\sin 2\theta
\bigr)
\end{equation}
with the normalization condition
\begin{equation}
    \int_{0}^{2\pi} D(f,\theta)\, d\theta = 1
\end{equation}
where $\lambda_k(f)$ are frequency-dependent Lagrange multipliers determined 
from the moment constraints derived from $\mathbf{G}(f)$.

\section{Experiments} \label{sec:expes}
We use the Bideford Bay buoy data obtained from UK National Network of Regional Coastal Monitoring Programmes (NNRCMP). 
These data are measured at the Bideford Bay station, England $(51^\circ 03.48' \,\mathrm{N}$, $004^\circ 16.62' \,\mathrm{W})$. 
The buoy is a Datawell Directional Waverider Mk III deployed at an approximate water depth of $11$m. 
We chose the 2024 raw buoy data and the buoy records three displacement components, 
including surface elevation (heave) and horizontal displacements in the north and west directions, 
with a sampling frequency of $1.28$Hz.\\

One-hour continuous data were used in this experiment. To evaluate reconstruction performance 
under different levels of data availability, random subsampling 
was performed at prescribed sampling ratios $r \in [0.1, 0.5]$, 
where $r$ denotes the fraction of observed samples retained from the full time series. 
Figure 1 shows an example of how the data are randomly sampled. 
In our computations, the mean value of each channel was removed before reconstruction 
to retain only information about the wave signal. A 99th percentile threshold was then applied 
via linear interpolation to mitigate the influence of spurious data points, 
and all channels were rescaled to the range $[-1,1]$.\\

\begin{figure}[t]
  \noindent\includegraphics[scale=0.9, angle=0]{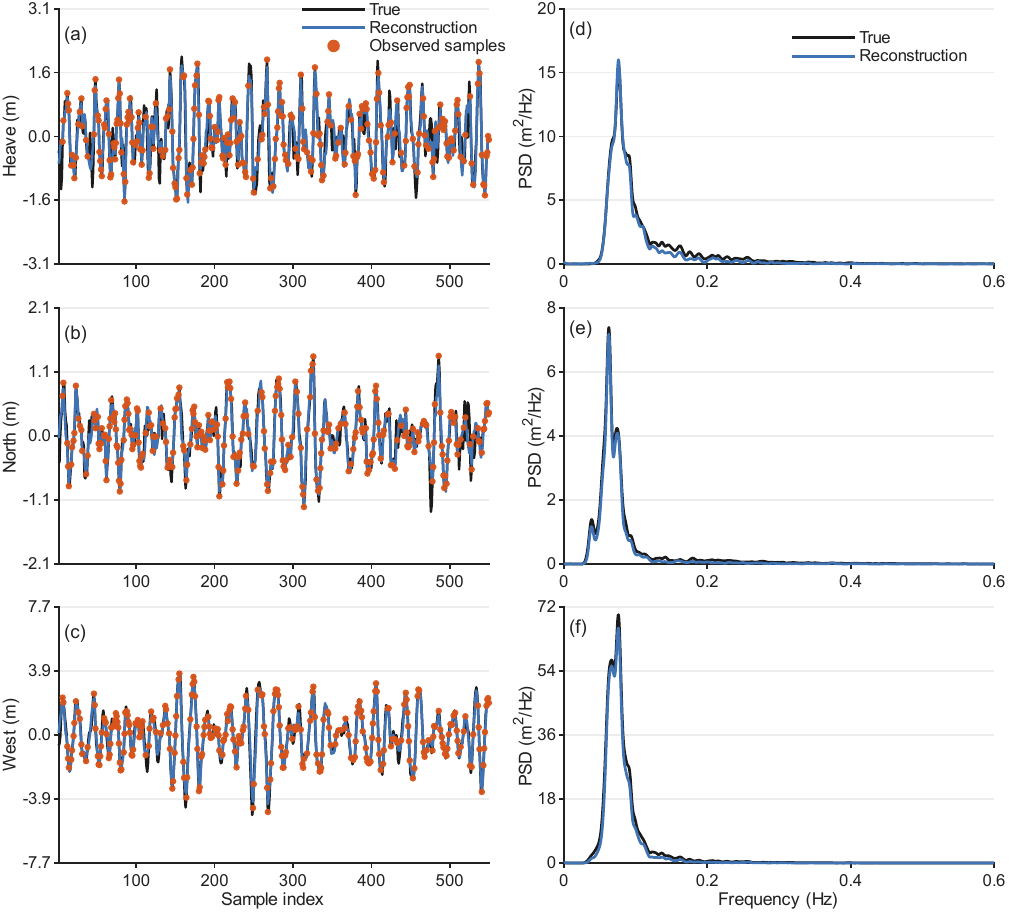}\\
  \caption{\small Illustration of random sampling and reconstruction performance at a 0.5 sampling ratio. Left column: time-domain signals for the heave, north, and west components, showing the true signal, sparse observations, and reconstructed results. Only 550 samples are displayed for visualization from the full record ($\approx$4600 samples). Right column: corresponding PSD comparisons, indicating that the dominant spectral peaks are accurately preserved despite substantial subsampling.}\label{f1}
\end{figure}

For the interpolation, each segment is $1$ hour of data and contains approximately $N\approx4600$ samples. 
Under this setting, we normalize each channel using the 99th percentile. 
Given a time series $x= \{x_1, x_2, ...,x_N\}$, the p-th percentile $x_{(p)}$ 
corresponds to the value below which $p\%$ the data fall. 
This is computed by sorting the data and applying linear interpolation
\begin{equation}
    x_{(p)} = x_{(i)} + \alpha \left( x_{(i+1)} - x_{(i)} \right),
\end{equation}
where $\mathrm{pos} = \frac{p}{100}(N - 1) + 1,\quad i = \lfloor \mathrm{pos} \rfloor,\quad \alpha = \mathrm{pos} - i$.
The normalized signal is then given by
\begin{equation}
    \tilde{x}(t) = \frac{x(t)}{x_{(99)}}
\end{equation}

Under this setting, $1\%\times N\approx 46$ samples which means that 
up to about $46$ extreme points can be tolerated without significantly affecting the normalization. 
The 99th percentile provides a robust estimate of the effective signal amplitude and is less sensitive to outliers.
We then apply our proposed model to the processed data and conduct experimental analysis 
by comparing performance of power spectral density (PSD) under different regularization methods, 
spectral conditions and further its impact on directional spectrum estimation under different methods.

\subsection{Reconstruction using Different Regularization Mechanism}

\begin{figure}[t]
\centering
  \noindent\includegraphics[scale=0.9, angle=0]{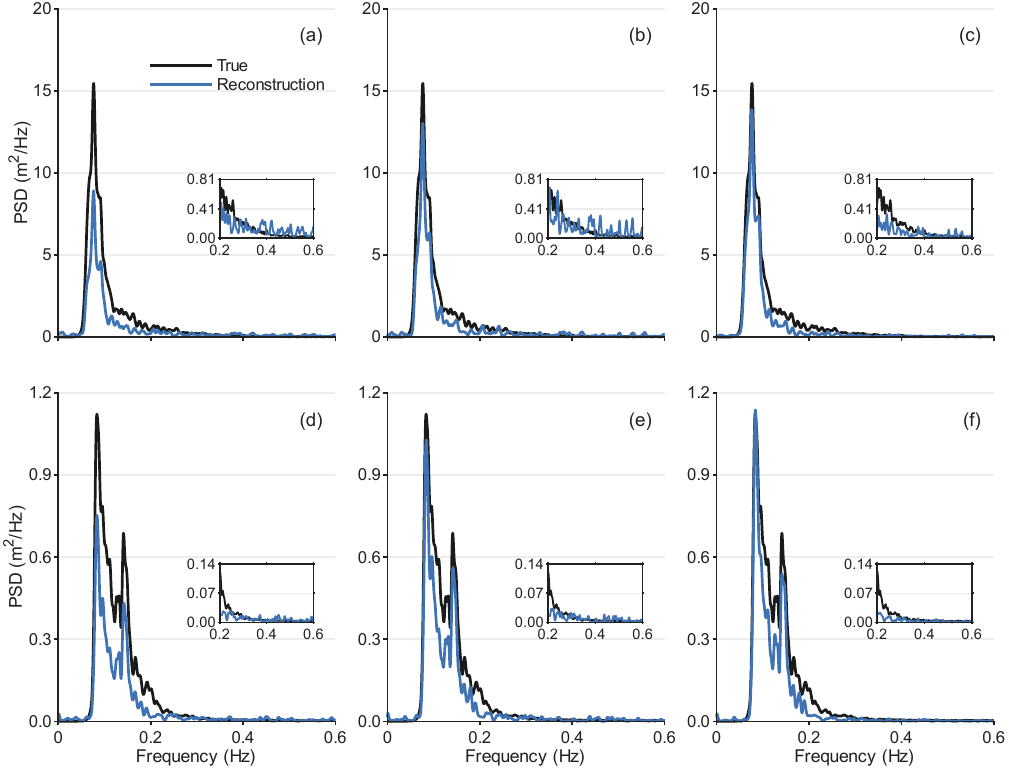}\\
  \caption{\small Comparison of reconstructed heave power spectral density (PSD) 
under different regularization strategies using random sampling 
(subsampling ratio $= 0.3$) 
with 1-hour continuous data. The top row shows a narrow spectrum case, 
while the bottom row corresponds to a bimodal spectrum with two dominant peaks. 
Panels (a) and (d) present reconstruction using element-wise sparsity ($\ell_1$ norm) only; 
(b) and (e) include an additional energy constraint; 
(c) and (f) show results from the full model incorporating element-wise sparsity, group sparsity, 
and energy correction. 
The black curve denotes the reference spectrum and the blue curve denotes the reconstruction. 
Insets highlight the high-frequency region.}\label{f1}
\end{figure}

Figure 2 compares the reconstructed heave spectra under $30 \%$ random sampling 
using different regularization strategies for both a narrow-band spectrum 
and a bimodal spectrum with two distinct wave systems. 
A clear improvement in reconstruction quality is observed as additional regularization components are introduced.\\

With element-wise sparsity alone (Figures 2a and 2d), the dominant spectral regions 
are correctly identified in both cases, indicating that the $\ell_1$ prior is sufficient to capture the main spectral support. 
This includes both the single peak in the narrow-band spectrum and the two dominant peaks in the bimodal case. 
However, the spectral amplitude is consistently underestimated across the frequency range, 
with the largest discrepancy occurring around the primary peaks. This reflects the shrinkage bias introduced by $\ell_1$ regularization.\\

With the energy constraint incorporated (Figures 2b and 2e), the overall 
amplitude level is significantly improved. The dominant peaks become much closer 
to the reference in magnitude, indicating the global energy level is effectively restored. 
As shown in the insets, deviations remain in the higher-frequency region, 
suggesting that amplitude correction alone does not fully regularize the spectral structure, 
particularly in the presence of multiple interacting components.\\

For the full model (Figures 2c and 2f), which combines element-wise sparsity, 
group sparsity, and energy correction, the reconstruction is further improved 
in both amplitude accuracy and structural consistency. 
The dominant peaks are accurately recovered in both location and magnitude, 
and the interaction between multiple spectral components is better preserved. 
In addition, the high-frequency deviations are substantially reduced, resulting in a smoother and more consistent spectral decay.\\

\subsection{Reconstruction under Different Spectral Conditions}
\begin{figure}[t]
\centering
  \noindent\includegraphics[scale=0.9, angle=0]{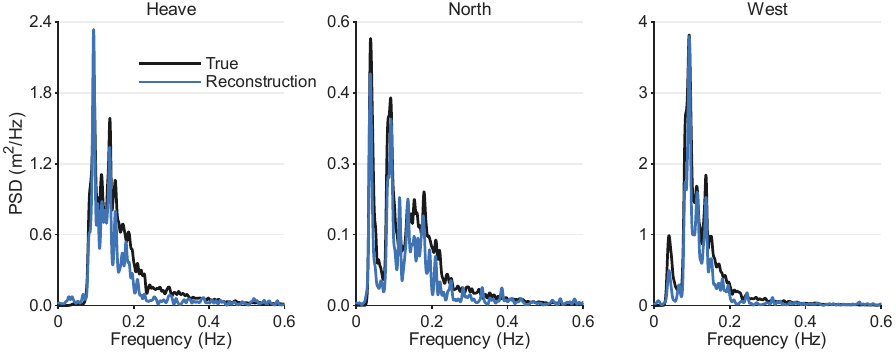}\\
  \caption{\small Reconstruction of power spectral density (PSD) for the three displacement components (heave, north, and west) 
using 1-hour continuous data with a sampling ratio of 0.35 for a relatively broad spectrum with a bimodal structure. 
The black curves denote the reference spectra and the blue curves denote the reconstructed spectra
using the full model incorporating element-wise sparsity, group sparsity, and energy correction. 
The two dominant peaks are consistently recovered across all channels, while the overall spectral shape shows 
increased spread compared to the narrow-band case. Differences among channels are evident 
in both amplitude and spectral decay, reflecting channel-dependent characteristics under broader spectral conditions.}\label{f1}
\end{figure}

Fig.~3 provides a representative example for the broader spectral case at a sampling ratio of 0.35. Compared with Figure 2, this experiment uses a higher sampling ratio. Although the three displacement components exhibit different spectral shapes, the main spectral features are recovered in all channels. The heave component captures the dominant low-frequency peak and subsequent decay, while the north and west components reproduce their respective low-frequency structures and overall spectral trends, with remaining differences mainly in weaker components.\\

\begin{figure}[t]
\centering
\noindent\includegraphics[width=0.65\textwidth]{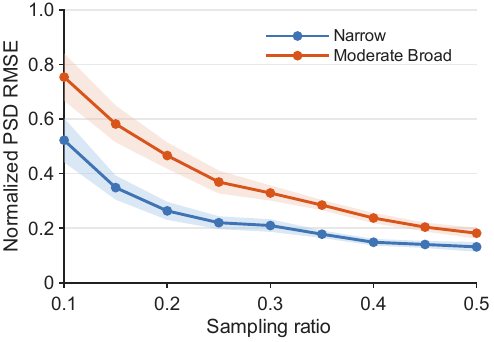}\\
\caption{Normalized PSD reconstruction error (RMSE) as a function of sampling ratio for narrow-band and moderately broad spectra. The solid lines denote the mean error over multiple random sampling realizations, and the shaded regions represent one standard deviation. The reconstruction error decreases as the sampling ratio increases for both spectral conditions. However, the moderately broad spectrum consistently exhibits higher error, indicating that spectral complexity significantly affects reconstruction accuracy. A transition region is observed around a sampling ratio of approximately 0.25, below which both the error level and variability increase more rapidly.}\label{f1}
\end{figure}

And then Fig.~4 is presented to show the normalized PSD reconstruction error as a function of sampling ratio for both narrow-band and broader spectral conditions. In both cases, the error decreases as the sampling ratio increases, with a rapid reduction at low sampling ratios followed by a more gradual decrease. A clear transition is observed around a sampling ratio of approximately 0.25, below which the error is higher and more variable, and above which the reconstruction becomes more stable. Across the full range, the broader spectrum consistently exhibits larger error than the narrow-band case. However, when the sampling ratio reaches 0.5, the difference between broadband and narrowband cases becomes less pronounced, as sufficient information is retained in both scenarios. As illustrated in Fig.~1, both cases can achieve high-quality reconstruction, with strong agreement in both the time domain and the frequency domain.\\

This difference reflects the fact that, as the spectral energy is distributed over a wider frequency range, more samples are required to achieve comparable reconstruction accuracy. But the RMSE of both bands will converge to around 0.2 due to the use of $\ell_1$-norm, as it will pull Fourier coefficients toward zero, producing a systematic shrinkage bias. Aside from the dominant frequencies, weak-energy components are often suppressed to zero, preventing the error from converging to zero.\\

Overall, these results show that broader spectra require a higher sampling ratio to maintain reconstruction quality, while increasing the sampling ratio enables consistent recovery of the main PSD features across all channels despite their different spectral characteristics.

\subsection{Impact on Directional Spectrum Estimation}
Figures 5 and 6 show the directional spectra estimated from the reconstructed signals using IMLM2 and MEM2 under the narrow-band and broader spectral conditions.\\

In the narrow-band case, both methods recover the dominant directional system accurately (Fig.~5b and Fig.~6b). The main energy is concentrated at the correct direction and frequency, and the peak location and intensity are well preserved. The reconstructed spectra closely follow the reference, with only minor deviations around the peak and weak spurious components at distant directions.\\

In the broader case, where two directional models are presented, both methods still capture the main structure of the spectrum (Fig.~5d and Fig.~6d). The dominant low-frequency system and the secondary directional component are both identified, and their relative positions are preserved. This indicates that the large-scale directional organization remains stable after reconstruction, even under increased spectral complexity.
\begin{figure}[t]
\centering
  \noindent\includegraphics[scale=0.8, angle=0]{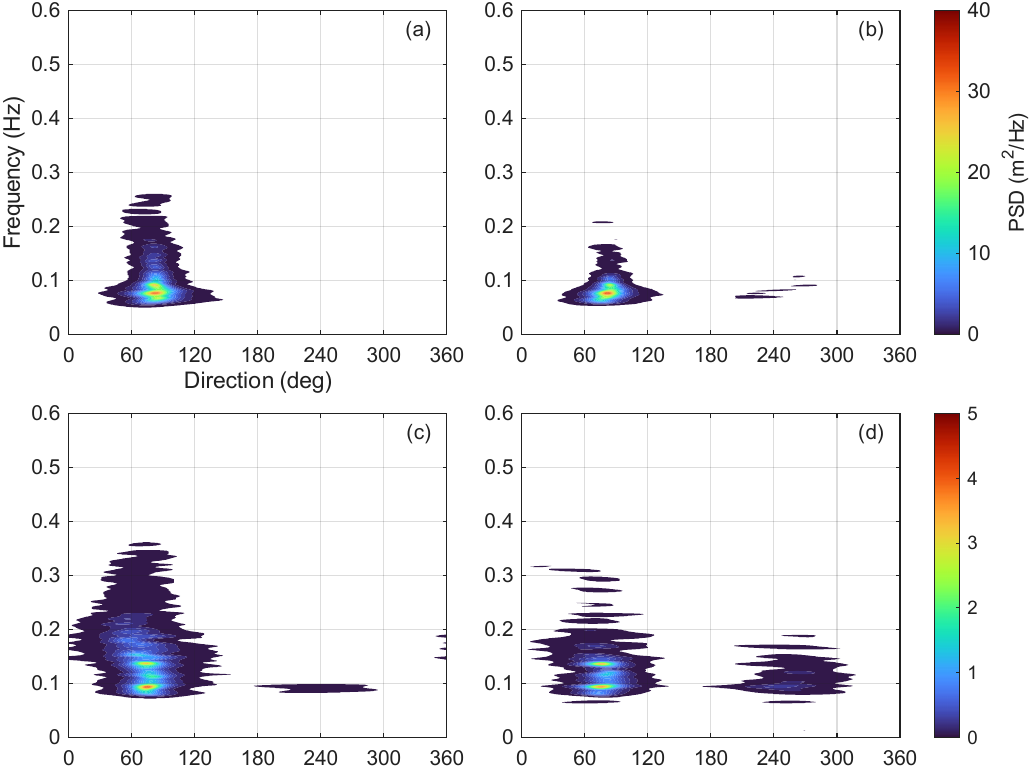}\\
  \caption{\small Directional wave spectra estimated using IMLM2. The top row corresponds to the narrow-band case derived from the signals shown in Figure 2, while the bottom row corresponds to a broader spectral condition obtained from the three reconstructed displacement components in Figure 4. Panels (a, c) show the reference spectra, and panels (b, d) show the reconstructed spectra. The sampling ratios are 0.3 for the top row and 0.35 for the bottom row. The dominant directional energy and overall structure are preserved after reconstruction.}\label{f1}
\end{figure}

\begin{figure}[t]
\centering
\noindent\includegraphics[scale=0.8, angle=0]{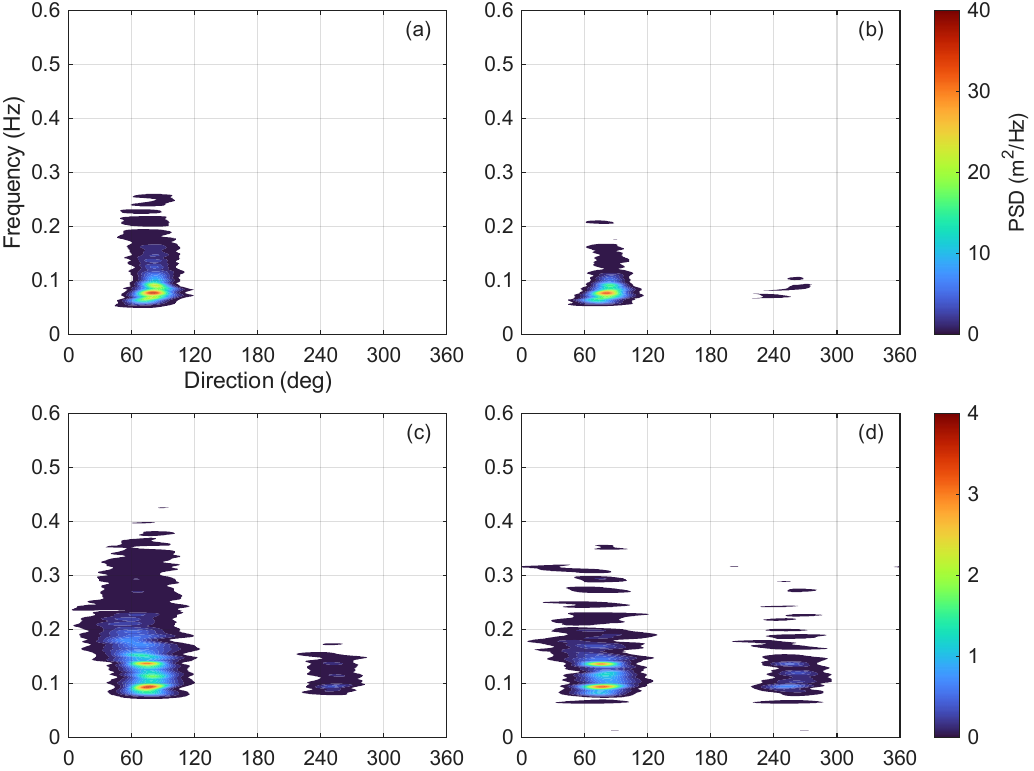}\\
\caption{\small Directional wave spectra estimated using MEM2. The top row corresponds to the narrow-band case derived from the signals shown in Figure 2, while the bottom row corresponds to the broader spectral condition obtained from the three reconstructed displacement components in Figure 4. Panels (a, c) show the reference spectra, and panels (b, d) show the reconstructed spectra. The sampling ratios are 0.3 for the top row and 0.35 for the bottom row. The dominant directional energy and overall structure are preserved, with a smoother distribution compared to the IMLM2 results.}\label{f1}
\end{figure}

\subsection{Error Analysis}
The error distributions shown in Figures 7 and 8 further support the reconstruction quality observed in Figures 5 and 6. To quantify the discrepancy, we define a signed error field over the frequency–direction domain. This formulation is consistent with standard signal analysis, where error is defined as the difference between reconstructed and reference signals. In line with image quality assessment studies, which emphasize structural differences rather than pointwise error magnitude \citep{wang2004image}, we further analyze the spatial distribution of the error:

\begin{equation}
    \varepsilon(f,\theta) = S_{\mathrm{rec}}(f,\theta) - S_{\mathrm{true}}(f,\theta)
\end{equation}
where $S_{\mathrm{rec}}(f,\theta)$ and $S_{\mathrm{true}}(f,\theta)$ denote the reconstructed and true directional spectra, respectively. To emphasize the spatial distribution of the discrepancy rather than its absolute scale, the error field is normalized by the peak value of the corresponding reference directional spectrum:

\begin{equation}
    \tilde{\varepsilon}(f,\theta)
    =
    \frac{\varepsilon(f,\theta)}
    {\max_{(f,\theta)} S_{\mathrm{true}}(f,\theta)}
\end{equation}

The domain frequency and direction are accurately captured, regardless of whether the spectrum is narrow-band or relatively broadband. A more detailed analysis of the error would require image similarity metrics, which are not considered in this study. Here, we present only the results obtained from the proposed method.\\

In Fig.~7 and Fig.~8, the deviations are not distributed as uniform background noise. Instead, they are largely confined to regions where the original spectra contain energy, indicating that compressed sensing maintains the global structure of the spectrum, while the discrepancies has localized and physically meaningful rather than random.\\

\begin{figure}[t]
\centering
  \noindent\includegraphics[scale=0.8, angle=0]{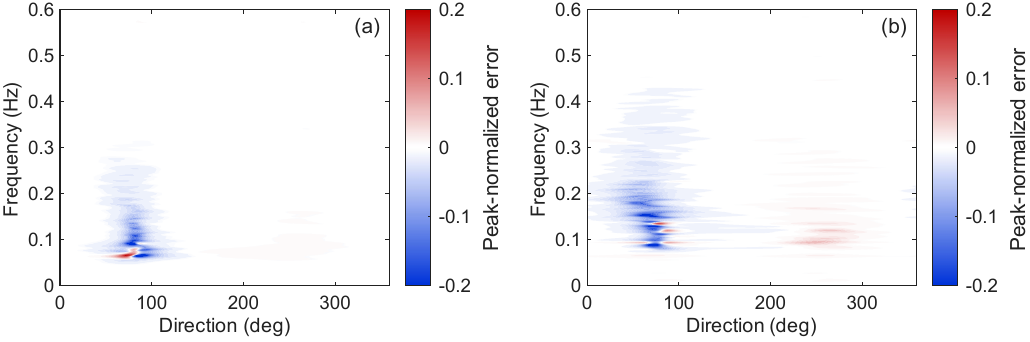}\\
  \caption{\small Peak-normalized signed error of the reconstructed directional spectra estimated using IMLM2. The error is calculated as the difference between the reconstructed and reference spectra, and is normalized by the peak value of the corresponding reference directional spectrum. Small error values below 0.01 of the maximum absolute error are masked to reduce visually negligible fluctuations. Panel (a) corresponds to the narrow-band case, while panel (b) corresponds to the broader spectral condition.}\label{f1}
\end{figure}
\begin{figure}[t]
\centering
  \noindent\includegraphics[scale=0.8, angle=0]{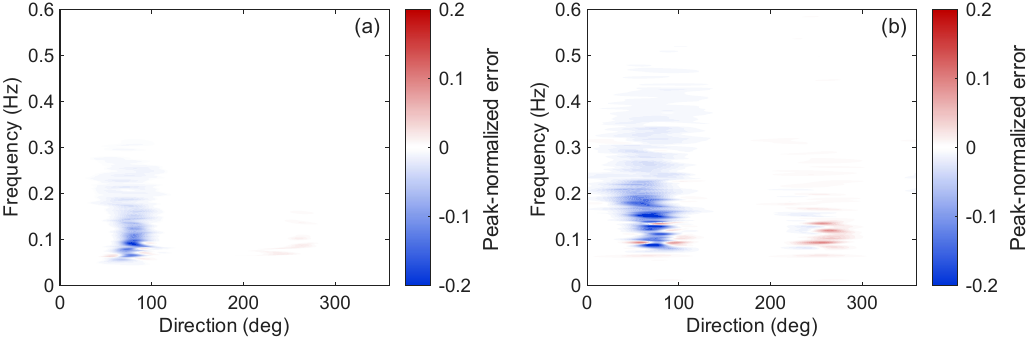}\\
  \caption{\small Peak-normalized signed error of the reconstructed directional spectra estimated using MEM2. The error is calculated as the difference between the reconstructed and reference spectra and is normalized by the peak value of the corresponding reference directional spectrum. Small error values below 0.01 of the maximum absolute error are masked to reduce visually negligible fluctuations. Panel (a) corresponds to the narrow-band case, while panel (b) corresponds to the broader spectral condition.}\label{f1}
\end{figure}

Building on this observation, a notable difference between IMLM2 and MEM2 can be found in how the reconstruction error is distributed. In the narrow-band case, IMLM2 shows a localized but more complex error structure around the dominant spectral region, with both overestimation and underestimation appearing near the main component (Fig.~7a). This indicates that the peak location is largely preserved, but the local energy distribution around the peak is not fully recovered. In contrast, MEM2 gives a smoother error pattern in the narrow-band case, where the error is mainly characterized by slight underestimation and no obvious spurious directional component (Fig.~8a).\\

As the spectrum broadens, the effective sparsity decreases and the limited sampling in compressed sensing reduces directional resolution, causing the error fields of both methods to spread over a wider frequency-direction range. For IMLM2, the broadband case shows a broad region of underestimation around the dominant directional sector, together with localized overestimation near the main component and weak spurious energy in the opposite direction (Fig.~7b). For MEM2, the error is also more widely distributed, and the opposite-direction spurious component becomes more distinct (Fig.~8b).\\

Overall, although compressed sensing successfully recovers the dominant directional structure, the remaining errors exhibit clear spatial organization rather than random behavior. The presence of coherent regions of overestimation and underestimation indicates a systematic redistribution of energy. This can be attributed to the fact that directional spectrum estimation relies on cross-spectral relationships among multiple channels, while compressed sensing inevitably degrades part of the inter-channel phase information during reconstruction. As a result, inconsistencies between channels are amplified in the cross-spectrum calculation, leading to structured errors in directional energy allocation rather than independent noise.\\

\section{Discussion and Limitations} \label{sec:discus}
\subsection{Sparsity, Spectral Structure, and Regularization Effects}
Independent sparsity modeling captures only intra-channel structure and fails to exploit shared support across channels, whereas multi-channel signals are often jointly sparse with a common support \citep{chen2014forest}. In our case, channel coupling is introduced by promoting joint sparsity across channels, which enforces a common support and improves reconstruction consistency \citep{eldar2009average}. This modeling choice is also physically motivated. Under linear wave theory, wave components at a given frequency are present simultaneously in all displacement channels, with consistent phase relationships determined by the wave direction \citep{dean1991water}. This constraint encourages frequency-wise co-existence and helps maintain the dominant spectral structure while reducing incoherent high-frequency noise. However, since the underlying conditions of compressed sensing, such as the RIP, are not strictly satisfied, the group constraint alone cannot guarantee accurate recovery.\\

As a consequence, $\ell_1$-based reconstruction introduces bias in the recovered coefficients due to its uniform shrinkage effect, which does not satisfy the unbiasedness property for large coefficients \citep{zou2006adaptive}. In the spectral domain, this leads to an underestimation of the total reconstructed energy. To address this issue, an additional energy term is introduced to compensate for the bias and restore the dominant spectral energy. In this paper, we derived a closed-form solution for the energy correction in the form of an isotropic rescaling, and its effectiveness therefore depends directly on the accuracy of the energy estimate. Nevertheless, the energy level is approximated from the sampled data and is not claimed to be optimal. A more accurate estimation of the energy would thus further improve the reconstruction.\\

\subsection{Sampling Strategy and Practical Limits}
In practical applications, when the underlying spectrum is not fully broadband and is not dominated by noise, moderate sampling ratios can be sufficient to recover the dominant spectral structure. This is consistent with the general principle that recovery performance in compressed sensing depends strongly on the sparsity or compressibility of the signal \citep{davenport2012introduction}. This is particularly evident in the directional spectrum, which remains stable even when the one-dimensional spectra of individual channels become less accurate under broadened conditions. Although such cases deviate from the ideal assumptions of compressed sensing, the dominant energy distribution can still be preserved in a statistical sense.\\

As the sampling ratio increases, the reconstruction accuracy improves consistently. At a sampling ratio of around 0.5, both the spectral representation and the time-domain signal can be recovered with high fidelity, as shown in Fig.~1. In this regime, the problem becomes less characteristic of compressed sensing and instead approaches a well-determined inverse problem with sufficient observations. For complex sea states with significantly broadened spectrum, a higher sampling ratio remains a preferable choice when stable reconstruction is required.\\

The sampling strategy itself can also be treated more flexibly. Instead of fixing the sampling ratio a priori, the problem can be viewed from a reverse perspective: the full signal can first be acquired, followed by a structure-aware compression step. By analyzing the spectral complexity over a given time window, the sampling density can then be selected accordingly. For relatively simple spectra with concentrated energy, lower sampling ratios (e.g., around 0.125 or even lower) can still provide accurate reconstruction. For moderately complex conditions, intermediate ratios such as 0.35 remain effective, while more complex spectra require higher sampling densities. Furthermore, the sampling patterns across channels need not be identical. In multi-channel acquisition settings, employing distinct sampling patterns across channels can improve recovery quality by providing more diverse measurements \citep{chun2017compressed}. This multi-channel sampling strategy works together with the adaptive selection of sampling ratios, forming a unified framework that accounts for both spectral structure and inter-channel correlations.\\

From this perspective, our proposed method can be interpreted as a direct subsampling approach, or as a dynamic compression framework. By combining structure-aware sampling ratios with channel-dependent sampling patterns, the method enables flexible data reduction while preserving the dominant spectral characteristics.

\subsection{Directional Spectrum Characteristics (IMLM2 vs MEM2)}
The differences between IMLM2 and MEM2 are mainly related to how each method responds to the reconstructed spectral input. In our framework, the directional estimation is performed on signals obtained from sparsity-based reconstruction, where the spectral energy is already relatively concentrated. Furthermore, when applied to reconstructed data, small inconsistencies in the input may propagate into the estimated spectrum. Spectral estimation procedures can be sensitive to errors in the underlying measurements and statistical estimates \citep{zhang2006new}, which can lead to the appearance of spurious peaks, particularly in low-energy regions.

\section{Conclusion} \label{sec:conclusion}
This paper presents a compressed sensing framework for reconstructing wave spectra from sparsely sampled multi-channel buoy data. By combining element-wise sparsity, cross-channel group structure, and an energy correction term, the method is able to recover the main spectral structure together with a consistent energy level under incomplete observations. In addition to noise, some structural issues remain, suggesting further consideration of physical regularization of the cross-spectrum matrix, or considering using joint analysis of two buoys instead of data from only one buoy. 

Construction the wave spectrum from sparse measurements could be important
in the future as more of the global wave climate is mapped out.
At present, the available information is still limited \citep{osorio2016construction}.
Looking forward, the method presented here could be combined with other known methods of enhancing accuracy of hindcasts and wave forecasts \citep{costa2023enhancing} or extract information from sparse data \citep{kuehn2023deep}. 
Incorporating 
further physics-based constraints may improve robustness under severe undersampling. Combining the compressed sensing
approach with adaptive or learning-based priors may allow the model capture evolving spectral features and enable
more accurate real-time applications such as already in use in short-time phase-resolved waveforecasting \citep{liu2022machine,holand2024real}.

In this context, the Fourier-domain formulation used in the present work is particularly natural, since surface wave records are inherently oscillatory signals and their energy is often concentrated around a limited number of dominant spectral components. Although the use of a discrete Fourier basis may introduce off-grid effects, such as spectral leakage and frequency drift, the results indicate that it remains a suitable and physically meaningful representation for sparse wave reconstruction. At the same time, the study shows that sparsity should not be interpreted too rigidly. The sparsest reconstruction is not necessarily the most physical one and may underestimate spectral energy, but with an appropriate energy correction it provides a strong modelling basis for recovering the wave spectrum with high fidelity.

%
%


%
%

\bibliography{QJreferences.bib}

\end{document}